\documentclass[aps,prl,twocolumn,superscriptaddress]{revtex4-1}

\usepackage{graphicx}
\usepackage{float}
\usepackage{dcolumn}
\usepackage{bm}
\usepackage{amsmath}
\usepackage{color}
\usepackage{verbatim}
\usepackage{natbib}
\usepackage{verbatim}
\usepackage[colorlinks,
            linkcolor=blue,
            anchorcolor=blue,
            citecolor=blue,
urlcolor=blue
            ]{hyperref}
\usepackage{color}
\usepackage{enumerate}

\newcommand{\RNum}[1]{\uppercase\expandafter{\romannumeral#1\relax}}

\usepackage{bm}

\begin{document}


\title{Resource-efficient Direct Characterization of General Density Matrix}


\author{Liang Xu}
\thanks{these authors contribute equally to this work}
\affiliation{College of Metrology and Measurement Engineering, China Jiliang University, Hangzhou, 310018, China}
\affiliation{National Laboratory of Solid State Microstructures, Key Laboratory of Intelligent Optical Sensing and Manipulation, College of Engineering and Applied Sciences, and Collaborative Innovation Center of Advanced Microstructures, Nanjing University, Nanjing 210093, China}
\affiliation{Research Center for Quantum Sensing, Zhejiang Lab, Hangzhou 310000, China}
\author{Mingti Zhou}
\thanks{these authors contribute equally to this work}
\affiliation{College of Metrology and Measurement Engineering, China Jiliang University, Hangzhou, 310018, China}
\affiliation{Research Center for Quantum Sensing, Zhejiang Lab, Hangzhou 310000, China}
\author{Runxia Tao}
\thanks{these authors contribute equally to this work}
\affiliation{College of Metrology and Measurement Engineering, China Jiliang University, Hangzhou, 310018, China}
\affiliation{Research Center for Quantum Sensing, Zhejiang Lab, Hangzhou 310000, China}
\author{Zhipeng Zhong}
\affiliation{Research Center for Quantum Sensing, Zhejiang Lab, Hangzhou 310000, China}
\author{Ben Wang}
\affiliation{National Laboratory of Solid State Microstructures, Key Laboratory of Intelligent Optical Sensing and Manipulation, College of Engineering and Applied Sciences, and Collaborative Innovation Center of Advanced Microstructures, Nanjing University, Nanjing 210093, China}
\author{Zhiyong Cao}
\affiliation{School of Electronic Control, Chang’an University, Xi’an 710064, China}
\author{Hongkuan Xia}
\affiliation{National Laboratory of Solid State Microstructures, Key Laboratory of Intelligent Optical Sensing and Manipulation, College of Engineering and Applied Sciences, and Collaborative Innovation Center of Advanced Microstructures, Nanjing University, Nanjing 210093, China}
\author{Qianyi Wang}
\affiliation{National Laboratory of Solid State Microstructures, Key Laboratory of Intelligent Optical Sensing and Manipulation, College of Engineering and Applied Sciences, and Collaborative Innovation Center of Advanced Microstructures, Nanjing University, Nanjing 210093, China}
\author{Hao Zhan}
\affiliation{National Laboratory of Solid State Microstructures, Key Laboratory of Intelligent Optical Sensing and Manipulation, College of Engineering and Applied Sciences, and Collaborative Innovation Center of Advanced Microstructures, Nanjing University, Nanjing 210093, China}
\author{Aonan Zhang}
\affiliation{Quantum Optics and Laser Science, Blackett Laboratory, Imperial College London, Prince Consort Rd, London SW7 2AZ, United Kingdom}
\author{Shang Yu}
\affiliation{Quantum Optics and Laser Science, Blackett Laboratory, Imperial College London, Prince Consort Rd, London SW7 2AZ, United Kingdom}
\author{Nanyang Xu}
\affiliation{Research Center for Quantum Sensing, Zhejiang Lab, Hangzhou 310000, China}
\author{Ying Dong}
\email{yingdong@cjlu.edu.cn}
\affiliation{College of Metrology and Measurement Engineering, China Jiliang University, Hangzhou, 310018, China}
\affiliation{Research Center for Quantum Sensing, Zhejiang Lab, Hangzhou 310000, China}
\author{Changliang Ren}
\email{renchangliang@hunnu.edu.cn}
\affiliation{Key Laboratory of Low-Dimensional Quantum Structures and Quantum Control of Ministry of Education,Key Laboratory for Matter Microstructure and Function of Hunan Province,
Department of Physics and Synergetic Innovation Center for Quantum Effects and Applications,Hunan Normal University, Changsha 410081, China}
\author{Lijian Zhang}
\email{lijian.zhang@nju.edu.cn}
\affiliation{National Laboratory of Solid State Microstructures, Key Laboratory of Intelligent Optical Sensing and Manipulation, College of Engineering and Applied Sciences, and Collaborative Innovation Center of Advanced Microstructures, Nanjing University, Nanjing 210093, China}


\date{\today}

\begin{abstract}
{Sequential weak measurements allow for the direct extraction of individual density-matrix elements, rather than relying on global reconstruction of the entire density matrix, which opens a new avenue for the characterization of quantum systems. Nevertheless, extending the sequential scheme to multi-qudit quantum systems is challenging due to the requirement of multiple coupling processes for each qudit and the lack of appropriate precision evaluation. To address these issues, we propose a resource-efficient scheme (RES) that directly characterizes the density matrix of general multi-qudit systems while optimizing measurements and establishing a feasible estimation analysis. In the RES, an efficient observable of the quantum system is constructed such that a single meter state coupled to each qudit is sufficient to extract the corresponding density-matrix element. An appropriate model based on the statistical distribution of errors is utilized to evaluate the precision and feasibility of the scheme. We have experimentally applied the RES to the direct characterization of general single-photon qutrit states and two-photon entangled states. The results show that the RES outperforms sequential schemes in terms of efficiency and precision in both weak- and strong-coupling scenarios. This work sheds new light on the practical characterization of large-scale quantum systems and the investigation of their non-classical properties.
}
\end{abstract}


\maketitle
\textit{Introduction.}--The density matrix provides a comprehensive description of a quantum system (QS), encompassing its intrinsic properties and interactions with other systems. Accurate characterization of the density matrix, particularly its off-diagonal elements, is crucial for revealing the non-classical properties of quantum systems \cite{horodecki2009quantum, modi2010unified, streltsov2017colloquium} and underpinning advanced quantum technologies \cite{degen2017quantum, braun2018quantum, giovannetti2011advances, vidrighin2014joint}. Quantum state tomography (QST) is a standard approach to globally reconstruct the whole density matrix based on the informationally-complete measurement results. However, the experimental challenges involved in implementing quantum measurements and the computational complexity of the reconstruction algorithm significantly increase with the size of the QS, rendering QST impractical for characterizing large-scale quantum states. Despite many efforts to improve the feasibility of QST \cite{wootters1989optimal, scott2010symmetric, bent2015experimental, banchi2018multiphoton, zhao2021fermionic}, characterizing the entire density matrix would be inefficient when only a few specific density-matrix elements are necessary to reveal certain properties, such as entanglement \cite{ friis2019entanglement} and coherence \cite{wu2021experimental}.

In recent years, weak measurement followed by post-selection has enabled the direct measurement of quantum wavefunctions for pure states \cite{lundeen2011direct, mirhosseini2014compressive, malik2014direct, fischbach2012quantum, bolduc2016direct, shi2015scan, pan2019direct}, as well as quasi-probability distributions of general mixed states \cite{lundeen2012procedure, salvail2013full, bamber2014observing, wu2013state}, which are equivalent to the density matrix through linear transformations \cite{dirac1945analogy}. Particularly, the density-matrix elements can be directly characterized through sequential measurements of complementary observables \cite{lundeen2012procedure, thekkadath2016direct, piacentini2016measuring, martinez2021theory}. Given the substantial benefits of reducing the number of measurement bases and avoiding the reconstruction algorithm, the direct-characterization schemes have been extended to quantum processes \cite{kim2018direct} and quantum detectors \cite{xu2021direct_PRL, xu2021direct_AP}. However, the direct characterization of the density matrix relies on sequential coupling processes for each qudit of the QS, which limits its efficiency and scalability for multi-particle and high-dimensional quantum states \cite{ren2019efficient}. Currently, the direct characterization of high-dimensional or two-photon density matrix typically requires introducing certain assumptions or utilizing extra resources \cite{zhou2021direct, chen2021directly}. Moreover, extracting high-order correlation information during sequential weak measurements results in large statistical errors \cite{maccone2014state, gross2015novelty}. To improve the precision of the direct characterization, a rigorous framework based on strong measurements has been developed \cite{zhu2016direct, vallone2016strong, calderaro2018direct, zhang2020direct}. However, this approach may destroy the unique advantages of weak measurement in slightly disturbing the quantum systems.

In this paper, we propose a resource-efficient scheme (RES) to directly characterize the density-matrix elements of general multi-qudit quantum systems. In the RES, an efficient observable is constructed through the Hadamard transformation, which enables to extract the density-matrix element of interest with a single coupling process for each qudit. We experimentally demonstrate the RES by directly characterizing the density matrix of single-photon qutrit states and two-photon entangled states during the unitary evolution and dephasing processes. Our results show that the RES achieves better precision than sequential schemes with fewer quantum measurements, irrespective of whether the coupling strength is weak or optimized in terms of precision.

\begin{figure}[t]
\centering
\includegraphics[width=0.45\textwidth]{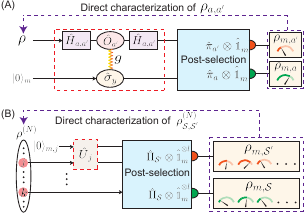}
\caption{The schematic diagram of the resource-efficient direct-characterization schemes for (A) single-qudit states and (B) multi-qudit states.
}\label{schematic}
\end{figure}

\textit{Theoretical framework.}--We begin by considering a general single-qudit state described by a $d$-dimensional density matrix $\rho = \sum_{a,a^\prime}\rho_{a,a^\prime}|a\rangle\langle a^\prime|$. To directly characterize the off-diagonal element $\rho_{a,a^\prime}$ ($a\neq a^\prime$), we associate it with the average value $\text{Tr}(\hat{M}_{a^\prime,a} \rho) = \rho_{a,a^\prime}$ of the operator $\hat{M}_{a^\prime,a} = d\hat{\pi}_{a^\prime}\hat{\pi}_b\hat{\pi}_a$, where $\hat{\pi}_b = |b\rangle\langle b|$ and $|b\rangle =\frac{1}{\sqrt{d}} \sum_a|a\rangle$. Typically, the observables $\hat{\pi}_a$ and $\hat{\pi}_b$ are sequentially measured using two independent meter states, followed by the post-selection projector $\hat{\pi}_{a^\prime}$ \cite{lundeen2012procedure, thekkadath2016direct, calderaro2018direct, martinez2021theory}. Joint measurements on the post-selected meter states lead to the average value $\text{Tr}(\hat{M}_{a^\prime,a} \rho)$. Alternatively, we can decompose $\hat{M}_{a^\prime,a} = \hat{\pi}_{a^\prime} \hat{C}_{a,a^\prime}$ into the product of an efficient observable $\hat{C}_{a,a^\prime} = |a\rangle\langle a^\prime| + |a^\prime\rangle\langle a| + \sum_{k\neq a, k\neq a^\prime} |k\rangle\langle k|$ and the post-selection projector $\hat{\pi}_{a^\prime}$. Consequently, by measuring the efficient observable, we can directly characterize the density-matrix element with only one coupling process for each qudit.

\begin{figure*}[htbp]
\centering
\includegraphics[width=0.95\textwidth]{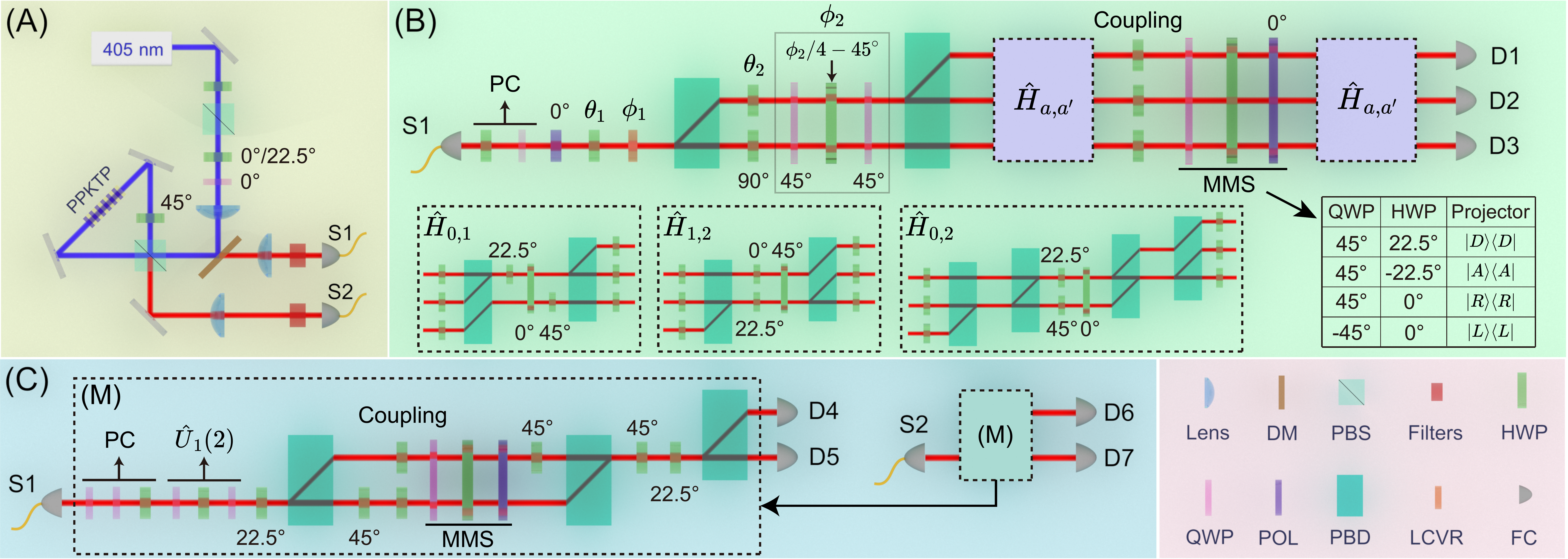}
\caption{Experimental setup. (A) A 405 nm continuous-wave (CW) laser is used to pump a periodically poled $\text{KTiOPO}_{\text{4}}$ (PPKTP) crystal inside a Sagnac interferometer for spontaneous parametric down-conversion (SPDC) to prepare heralded single-photon or two-photon polarization-entangled states. (B) The single-photon path-encoded qutrit states are prepared and characterized directly. (C) The two-photon polarization-entangled states are characterized directly by separately inputting two photons into the corresponding setup. The measurements of the meter states (MMS) are realized with the angles of wave plates listed. PC, polarization compensation; DM, dichroic mirror; PBS, polarizing beam splitter; POL, polarizer; LCVR, liquid-crystal variable retarder; FC, fiber coupler.
}\label{Expsetup}
\end{figure*}

Fig. \ref{schematic} (A) depicts the RES for directly characterizing the density matrix of a single-qudit QS. Within the red dashed box, the efficient observable $\hat{C}_{a,a^\prime}$ is constructed via the Hadamard transformation $\hat{H}_{a,a^\prime}$ on the ordinary observable $\hat{O}_{a^\prime}=\hat{\openone}_s - 2|a^\prime\rangle\langle a^\prime|$ of the QS, given by $\hat{C}_{a,a^\prime} = \hat{H}_{a,a^\prime}\hat{O}_{a^\prime}\hat{H}_{a,a^\prime}$. A single-qubit meter state (MS) $|0\rangle_m$ that can be either another particle or another degree of freedom of the same particle is coupled with the QS $\rho$ via the Hamiltonian $\hat{H} = G(t)\hat{C}_{a,a^\prime}\otimes \hat{\sigma}_y$, where $\hat{\sigma}_y$ is the Pauli operator of the MS. We define the coupling strength as $g = \int G(t) dt$ with $G(t)$ a compact support around the time of coupling. After the coupling process $\hat{U}=\exp(-i\int \hat{H} dt)$, the system-meter joint state evolves to $\rho_{jt} = \hat{U} \rho \otimes |0\rangle_m\langle 0| \hat{U}^\dagger$. Given a pair of post-selection projectors $\hat{\pi}_{a^\prime}$ and $\hat{\pi}_a$ corresponding to the conjugate operators $\hat{M}_{a^\prime,a}$ and $\hat{M}_{a,a^\prime}$, we perform both post-selection on the QS and measure the observables $\hat{\sigma}_+ = \hat{\sigma}_x+i\hat{\sigma}_y$ and $\hat{\sigma}_- =  \hat{\sigma}_x - i\hat{\sigma}_y$ of the MS to obtain the weak average $\text{Tr}(\hat{M}_{a^\prime,a}\rho)$ and the complex conjugate of $\text{Tr}(\hat{M}_{a,a^\prime}\rho)$, respectively. Thus, the density-matrix element can be derived by
\begin{equation}
\rho_{a,a^\prime} = \frac{1}{2\sin(2g)}\text{Tr} \big[(\hat{\Pi}_{a^\prime,+} + \hat{\Pi}_{a,-})  \rho_{jt}\big],
\end{equation}
where $\hat{\Pi}_{a^\prime,+} = \hat{\pi}_{a^\prime} \otimes \hat{\sigma}_+$ and $\hat{\Pi}_{a,-} = \hat{\pi}_{a} \otimes \hat{\sigma}_-$ denote the joint observables of the QS and the MS.

In Fig. \ref{schematic} (B), our formalism is extended to the direct characterization of $N$-qudit density matrix $\rho^{(N)} = \sum_{\mathcal{S},\mathcal{S}^\prime} \rho^{(N)}_{\mathcal{S},\mathcal{S}^\prime}|\mathcal{S}\rangle\langle \mathcal{S}^\prime|$, where $|\mathcal{S}\rangle\langle \mathcal{S}^\prime| = |a_1,...,a_N\rangle\langle a_1^\prime,...,a_N^\prime|$ and $|a_n\rangle$ represents the eigenstate of the $n$th qudit. To directly characterize the off-diagonal element $\rho_{\mathcal{S},\mathcal{S}^\prime}$, we decompose the operator $\hat{M}_{\mathcal{S}^\prime,\mathcal{S}} = |\mathcal{S}^\prime\rangle\langle \mathcal{S}|$ (or its conjugate $\hat{M}_{\mathcal{S},\mathcal{S}^\prime}$) into the product of the joint efficient observable $\hat{C}^{(N)} = \otimes_{n=1}^{N}\hat{C}_n$ and the post-selection projector $\hat{\Pi}_{\mathcal{S}^\prime}^{(N)} = |\mathcal{S}^\prime\rangle\langle \mathcal{S}^\prime|$ ($\hat{\Pi}_{\mathcal{S}}^{(N)}$). The efficient observable $\hat{C}_n$ for the $n$th qudit is divided into the following two cases: (i) if $a_j \neq a_j^\prime$, we choose $\hat{C}_j = |a_j\rangle \langle a_j^\prime| +|a_j^\prime\rangle \langle a_j| + \sum_{k_j\neq a_j, k_j\neq a_j^\prime} |k_j\rangle\langle k_j| $, and the coupling process is analogous to that in the single-qudit situation; (ii) if $a_k = a_k^\prime$, $\hat{C}_k$ is equal to $\hat{\openone}_{s,k}$, and no coupling is required. The separate measurement of the efficient observable $\hat{C}_j$ on each qudit with the identical coupling strength $g$ is implemented by the Hamiltonian $\hat{H}^{(N)}= G(t)\sum_{a_j \neq a_j^\prime}\hat{C}_j\otimes \hat{\sigma}_{y,j}$. Assuming that there are totally $l$ independent couplings in the $N$-qudit system, we get the joint state $\rho_{jt}^{(N)} = \hat{U}^{(N)} \Big[ \rho^{(N)}\otimes (|0\rangle_m\langle 0|)^{\otimes l} \Big]  \hat{U}^{(N)\dagger}$ with the unitary operator $\hat{U}^{(N)} = \exp[-i\int \hat{H}^{(N)} dt]$. The direct product of the post-selection operator $\hat{\Pi}_{\mathcal{S}^\prime}$ ($\hat{\Pi}_{\mathcal{S}}$) and the observable $\otimes_{j=1}^l \hat{\sigma}_{+,j}$ ($\otimes_{j=1}^l \hat{\sigma}_{-,j}$) leads to the joint measurement operator $\hat{\Pi}_{\mathcal{S}^\prime,+}^{(l)}$ ($\hat{\Pi}_{\mathcal{S},-}^{(l)}$). Thus, the density-matrix element can be directly obtained by 
\begin{equation}
\rho_{\mathcal{S},\mathcal{S}^\prime} = \frac{1}{2\sin^l (2g)} \text{Tr}\big[ (\hat{\Pi}_{\mathcal{S}^\prime,+}^{(l)} + \hat{\Pi}_{\mathcal{S},-}^{(l)} )\rho_{jt}^{(N)}   \big].
\end{equation}

To comprehensively evaluate the precision of the direct-characterization schemes, we analyze the statistical errors of the measured density-matrix elements. Given that the quantum states are uniformly sampled over the state space, the statistical properties of errors provide an insight into the expected precision and feasibility of the scheme. Specifically, the arbitrary qudit state $\rho_d = \hat{V} |0\rangle\langle 0| \hat{V}^\dagger$ is sampled by evolving the initial state $|0\rangle\langle 0| = (1,0,...,0)^T(1,0,...,0)$ under the unitary operator $\hat{V}\in \hat{U}(d)$. We denote the variances of the real and imaginary parts of $\rho_{a,a^\prime} (a\neq a^\prime)$ as $\delta^2 \text{Re}(\rho_{a,a^\prime})$ and $\delta^2 \text{Im}(\rho_{a,a^\prime})$, respectively. With the Haar measure $\mu_d(\hat{V})$, the variance $\delta^2 \text{Re}(\rho_{a,a^\prime})$ or $\delta^2 \text{Im}(\rho_{a,a^\prime})$ for any $a\neq a^\prime$ can be equivalently represented by $\delta^2_d$ to quantify the mean precision of single-qudit characterization as 
\begin{equation}
\Delta^2_d = \int _{\hat{V}\in \hat{U}(d)} \delta^2_d \text{d} \mu_d(\hat{V}).
\end{equation}
For $N$-qudit quantum systems, we evolve the $n$th qudit from a maximally entangled state $|\Psi_0\rangle = 1/\sqrt{d}\sum_{m=0}^{d-1} |m,...,m\rangle$ with the unitary operator $\hat{V}_n\in \hat{U}(d)$ to obtain an arbitrary sampling state $|\Psi_E\rangle = \otimes_{n=1}^N \hat{V}_n |\Psi_0\rangle$. We focus on the completely off-diagonal density-matrix elements $\rho^{(N)}_{\mathcal{S},\mathcal{S}^\prime}$ which satisfy $a_n \neq a_n^\prime$ for all $n$ since such elements imply the coherence information among all qudits. With the Haar measure $\mu_d(\hat{V}_n)$ for all the $n$th qudits, the variance of the real or imaginary part of an arbitrary completely off-diagonal density-matrix element, i.e., $\delta^2 \text{Re}[\rho^{(N)}_{\mathcal{S},\mathcal{S}^\prime}]$ or $\delta^2 \text{Im}[\rho^{(N)}_{\mathcal{S},\mathcal{S}^\prime}]$, can be equivalently represented by $\delta^2_{N,d}$ to quantify the mean precision of $N$-qudit characterization as
\begin{equation}
\Delta^2_{N,d} = \idotsint_{\hat{V}_n \in \hat{U}_n(d)} \delta^2_{N,d} \prod_{n=1}^N \text{d} \mu_{d}(\hat{V}_n).
\end{equation}
The equivalent variances $\delta^2_d$ and $\delta^2_{N,d}$ that depend on the specific quantum measurements are theoretically derived in Supplementary materials \cite{ODST_supplementary}.

\begin{figure*}[htbp]
\centering
\includegraphics[width=0.95\textwidth]{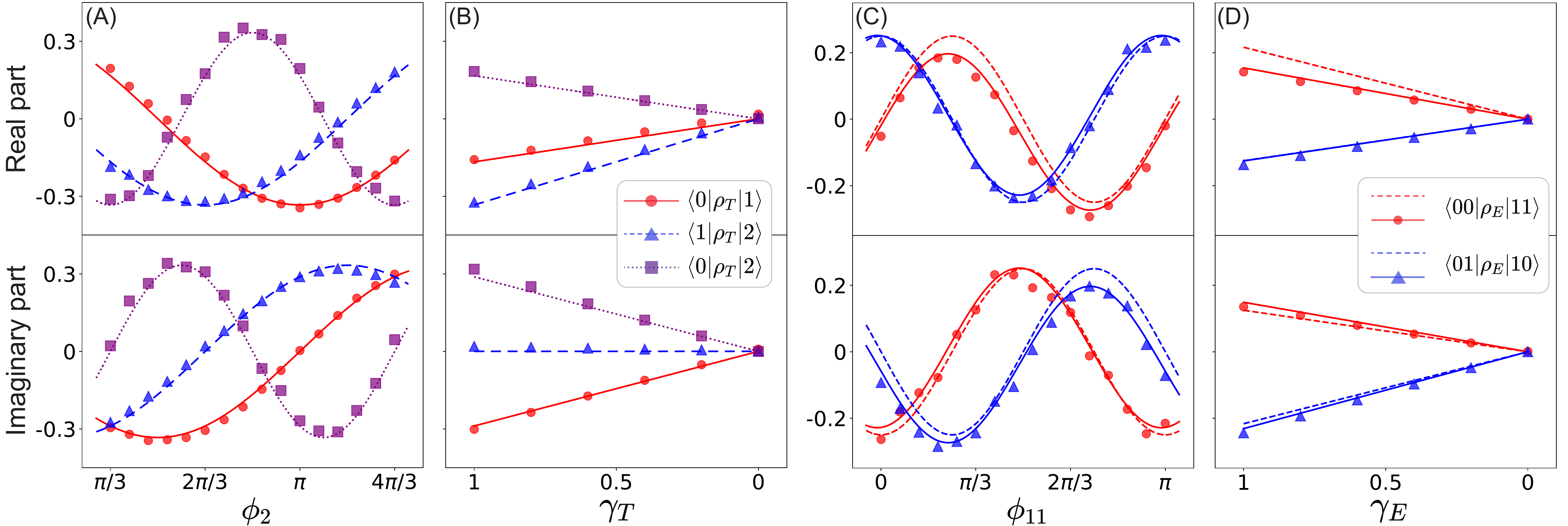}
\caption{The density-matrix elements of the qutrit states $\rho_T$ and two-photon entangled states $\rho_E$ are experimentally measured during the unitary and dephasing processes, with the correspondence: (A) $\rho_T$, unitary; (B) $\rho_T$, dephasing; (C) $\rho_E$, unitary; and (D) $\rho_E$, dephasing. In (A) and (B), solid (circular), dashed (triangular), and dotted (quadrate) lines (points) are used to represent theoretical (experimental) results. In (C) and (D), dashed lines represent ideal theoretical results while solid lines are derived according to realistic states. Circular and triangular points are used to depict experimental results.}
\label{Expresults}
\end{figure*}

\begin{figure}[hb]
\centering
\includegraphics[width=0.45\textwidth]{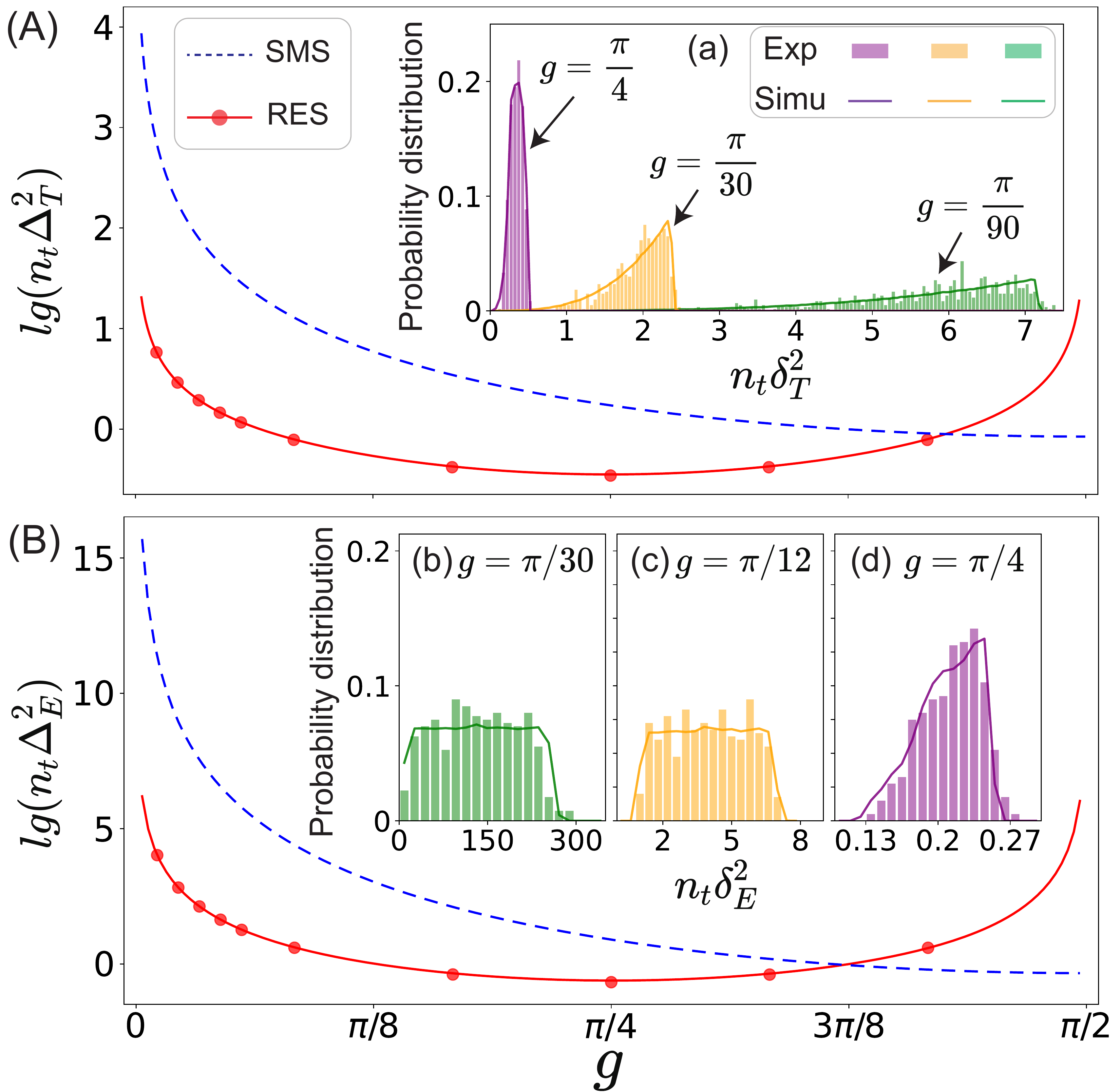}
\caption{Comparisons of mean precision between sequential measurement schemes (SMS) and the resource-efficient scheme (RES) are made in (A) and (B) for qutrit states and two-photon entangled states, respectively. The subscripts $T$ and $E$ denote the qutrit states and two-photon entangled states, respectively. Insets (a) to (d) illustrate the distribution of statistical errors using a bar chart, with corresponding lines obtained from numerical simulations of $10^4$ states.
}\label{Precision}
\end{figure}

\textit{Experiment.}--In order to demonstrate the feasibility of the RES, we experimentally apply it to the direct characterization of several typical photonic quantum states. The experimental setup for the preparation and direct characterization of single-qutrit states $\rho_T$ is shown in Fig. \ref{Expsetup} (B). Heralded single photons generated in module (A) input S1 port and subsequently pass through a half-wave plate (HWP), a quarter-wave plate (QWP), and a polarizer to initialize the polarization to the horizontal state $|H\rangle$. We employ two HWPs positioned at angles of $\theta_1$ and $\theta_2$, respectively, for polarization rotation. The relative phases $\phi_1$ and $\phi_2$ between the horizontal and vertical polarization are adjusted by a liquid-crystal variable retarder (LCVR, Thorlabs LCC1423-B) and a QWP-HWP-QWP combination, respectively. The polarizing beam displacers (PBDs) implement beam splitting by transmitting (refracting) photons in $|V\rangle$ ($|H\rangle$). With the optical paths $\{\text{up}, \text{middle}, \text{bottom}\}$ corresponding to the basis $\{|0\rangle, |1\rangle, |2\rangle\}$, we prepare an arbitrary pure qutrit state $|\psi_T\rangle =  \cos 2\theta_1 \cos 2\theta_2 |0\rangle + \cos2\theta_1\sin 2\theta_2e^{i\phi_2} |1\rangle + \sin 2\theta_1e^{i(\phi_1+\phi_2)} |2\rangle$ \cite{ODST_supplementary}. Mixed qutrit states are obtained by probabilistically mixing specific pure states \cite{ODST_supplementary}. 

We realize the Hadamard transformations ``$\hat{H}_{0,1}$", ``$\hat{H}_{1,2}$" and ``$\hat{H}_{0,2}$" on the qutrit states with three steps: (i) the first three HWPs and the following PBD transform the path-encoded information to the polarization; (ii) the middle three HWPs perform the Hadamard transformation on the polarization; (iii) the last three HWPs recover the polarization of photons at all paths to $|H\rangle$, initializing the MS \cite{ODST_supplementary}. To realize the coupling operation, the HWPs at all paths are rotated by $g/2$ degrees, except for the HWP at the path $a^\prime$, which is rotated by $-g/2$ degrees. The measurements of the meter states (MMS) are performed using a QWP-HWP-polarizer combination. After the surviving photons undergo the second Hadamard transformation, the photons are detected by the detectors D1, D2 and D3, corresponding to the successful post-selection by the operators $\hat{\pi}_0$, $\hat{\pi}_1$ and $\hat{\pi}_2$, respectively.

As shown in Fig. \ref{Expsetup} (C), two-photon entangled states $\rho_E$ are directly characterized by separately inputting two photons into S1 and S2 ports, respectively. Both photons undergo identical setups illustrated in the dotted box labeled ``(M)"  with the specific configurations depending on the density-matrix element of interest. We refer to the polarization degree of freedom of photons as the QS. Photons first get through a QWP-QWP-HWP combination for compensating the polarization changes in fiber transmission \cite{shao2007realization}. Then, we implement an arbitrary unitary transformation $\hat{U}_{1}(2)$ on the QS with a QWP-HWP-QWP combination \cite{simon1989universal, simon1990minimal}. The subsequent HWP (as well as the last HWP) at $22.5^\circ$ implements the Hadamard transformation on the polarization qubit. To facilitate the coupling between the QS and the MS, we transform the polarization information of photons to the path information through a PBD and the polarization of photons at both paths is initialized to $|H\rangle$ as the MS. The coupling operation and MMS are analogous to the single-qutrit experiment. We capture the coincidental counts by detecting the simultaneous arrival of two photons at specific pairs of detectors including $\text{D4} \& \text{D6}$, $\text{D4} \& \text{D7}$, $\text{D5} \& \text{D6}$ and $\text{D5} \& \text{D7}$ to realize the post-selection operators $\hat{\Pi}_{00}^{(2)}$, $\hat{\Pi}_{01}^{(2)}$, $\hat{\Pi}_{10}^{(2)}$ and $\hat{\Pi}_{11}^{(2)}$, respectively.

\textit{Results.}--In Fig. \ref{Expresults} (A), we depict the experimental results of directly characterizing the density-matrix elements of qutrit states during the unitary process by setting the parameters $\theta_1 = \arcsin(\sqrt{1/3})/2$, $\theta_2 = \pi/8$, $\phi_1 = \phi_2 + \pi/3$ and varying $\phi_2$ from $\pi/3$ to $4\pi/3$. Furthermore, we select a qutrit state $\rho_{T,0}$ by setting $\phi_2 = 2\pi/3$ and subject it to a dephasing process $\mathcal{N}(|a\rangle\langle a^\prime|) = \gamma_T |a\rangle\langle a^\prime |$, in which the dephasing coefficient ($0\le \gamma_T \le 1$ is adjusted by mixing specific pure states with different probabilities \cite{ODST_supplementary}. The measured results of the decreased off-diagonal elements $\gamma_T \langle a|\rho_{T,0}|a^\prime\rangle$ are shown in Fig. \ref{Expresults} (B). To directly characterize two-qubit systems, we evolve one of the Bell states $|\psi_b\rangle = (|H\rangle|V\rangle + |V\rangle|H\rangle) /\sqrt{2}$ under the composed unitary $\hat{U}_1(2)\otimes \hat{U}_2(2)$ with $\hat{U}_n(2) = \begin{pmatrix} \cos 2\theta_n e^{-i\phi_{n1}}                     & -\sin 2\theta_n e^{i\phi_{n2}} \\ \sin 2\theta_n e^{-i\phi_{n2}} & \cos 2\theta_n e^{i\phi_{n1}} \end{pmatrix}$ to prepare the sampling state $|\psi_E\rangle = \hat{U}_1(2)\otimes \hat{U}_2(2)|\psi_b\rangle$ and its density matrix $\rho_E = |\psi_E\rangle\langle \psi_E|$. By setting the parameters $\theta_{11} = \pi/8$, $\theta_{21}=\phi_{21}=\phi_{22} = 0$, $\phi_{12} = \phi_{11}+\pi/4$ and varying $\phi_{11}$ from $0$ to $\pi$, we illustrate the completely off-diagonal density-matrix elements, i.e., $\langle 00|\rho_E|11\rangle$ and $\langle 01|\rho_E|10\rangle$, in Fig. \ref{Expresults} (C). Next, an original state $\rho_{E,0} = |\psi_{E,0}\rangle\langle \psi_{E,0}|$ with $\phi_{11}=\pi/3$ undergoes a dephasing process $\mathcal{N}(|\mathcal{S}\rangle\langle \mathcal{S}^\prime|) = \gamma_E |\mathcal{S}\rangle\langle \mathcal{S}^\prime|$ with the coefficient $\gamma_E$ adjusted by appropriately mixing certain unitary operators $\hat{U}_1(2)$ and $\hat{U}_2(2)$ \cite{ODST_supplementary}. The experimental completely off-diagonal density-matrix elements of $\rho_E$ during the dephasing process are shown in Fig. \ref{Expresults} (D).

In Fig. \ref{Precision}, we compare the precision of the RES and sequential schemes with different coupling strength $g$. Provided that the total number of single photons or photon pairs per unit time follows a Poisson distribution with an average value of $n_t$, the mean characterization precision that is achieved with a single copy of single-qutrit and two-qubit states are theoretically derived as $n_t\Delta^2_T$ and $n_t\Delta^2_E$, respectively \cite{ODST_supplementary}. In the experiment, we randomly prepare 100 single-qutrit states $|\psi_T\rangle = \hat{U}(3)|0\rangle$ or two-photon polarization states $|\psi_E\rangle = \hat{U}_1(2)\otimes \hat{U}_2(2)|\psi_b\rangle$ with the unitary operators $\hat{U}(3)$, $\hat{U}_1(2)$ and $\hat{U}_2(2)$ sampled according to the Haar measure. The experimental precision is obtained from Monte Carlo simulation based on raw data. The results show that the precision of both direct-characterization schemes is improved with stronger coupling strength $g$. The optimal precision of the RES in characterizing both single-qutrit states ($g=\pi/4$, $n_t\Delta^2_T = 0.125$) and two-qubit states ($g=\pi/4$, $n_t\Delta^2_E = 0.208$) is overall better than that of sequential schemes with single-qutrit states ($g=\pi/2$, $n_t\Delta^2_T = 0.708$) and two-qubit states ($g=\pi/2$, $n_t\Delta^2_E = 0.458$).

\textit{Discussion and conclusions.}--The insets (a) to (d) of Fig. \ref{Precision} show non-divergent statistical errors, which confirms the feasibility of our scheme for characterizing arbitrary quantum states. In weak-measurement scenario (i.e., $g\rightarrow 0$), the precision of the RES is significantly better than that of sequential schemes with a ratio following $g^{2N}$ for $N$-qudit systems. This advantage makes our scheme more suitable for studying weak-measurement problems, such as the share of non-locality \cite{hu2018observation}, the error-disturbance relationship \cite{mao2019error}, and so on \cite{blok2014manipulating, kim2012protecting}. With qubit meter states used for the direct characterization of $N$-qudit systems, the resource-efficient or sequential schemes typically perform $2^{2N}$ or $2^{3N}$ projective measurements, respectively. When the efficiency of direct characterization schemes is compared quantitatively by counting the total number of photons (or equivalent time consumption) required to achieve the same precision, the mean efficiency of the optimal RES is 11.3 or 8.8 times higher than that of the optimized sequential schemes for single-qutrit states or two-qubit states, respectively.

In conclusion, we have proposed a resource-efficient scheme (RES) that can be used to directly characterize the density matrix of general multi-qudit quantum systems. Compared to sequential schemes, the RES only requires one coupling process for each qudit to extract the arbitrary density-matrix element of interest through an efficient observable. We experimentally demonstrate the advantages of the RES over sequential schemes by directly characterizing qutrit and two-qubit states with fewer measurements and better precision. These advantages are significantly enhanced as the particle number increases in weak coupling scenarios. Even when both the resource-efficient and sequential schemes are optimized in terms of precision over the coupling strength, the former still demonstrates an approximately tenfold efficiency advantage over the latter. Our work provides a promising approach for the characterization and exploration of large-scale quantum systems, with broad extensions from integrated photonic chip to other physical systems.

\begin{acknowledgments}
This work was supported by China Postdoctoral Science Foundation (Grants No. 2022M722918, No. 2021M702976, No. 2020TQ0297 and No. 2020M681949), International Postdoctoral Exchange Fellowship Program 2022 (Grant No. PC2022049), National Natural Science Foundation of China (Grants No. 12204437, No. 61975077, No. 12075245,  No. 12247105, No. 92265114 No. 12204434, and No. 12305034), the Center initiated Research Project of Zhejiang Lab (Grant No. 2021MB0AL01), the Natural Science Foundation of Hunan Province (Grant No. 2021JJ10033), Xiaoxiang Scholars Project of Hunan Normal university, the National Key Research and Development Program of China (Grant No. 2019YFA0308700) and Fundamental Research Funds for the Central Universities (Grant No. 020414380175) and and Zhejiang Provincial Natural Science Foundation of China (Grants No. LZ24A050006 and No. LQ24A040012).
\end{acknowledgments}

\end{document}